\documentstyle[preprint]{jpsj}

\input epsf

\def\runtitle{Comparative review of aging properties in spin glasses
and other disordered materials} \inst { Service
de Physique de l'Etat Condens\'e, CEA Saclay, 91191 Gif sur Yvette
Cedex, France } 

\recdate { \today } 

\abst { 
Aging phenomena have been studied in very different materials like polymers,
supercooled liquids or disordered orientational crystals. We recall here the main
features of aging in spin glasses, and use this example of magnetic
systems as a guideline for the description of the others. A particular
attention is put on the sensitivity of aging to the cooling rate and
to temperature variations. This allows us to
point out differences between {\it temperature specific} processes,
yielding ``rejuvenation and memory effects'' as known in
spin glasses, and {\it domain growth} processes, giving cumulative
contributions at different temperatures. The relevance of wall
depinning processes to rejuvenation and memory
effects  is
discussed at the light of recent results on disordered ferromagnets.
} 

\kword { aging,
glassy dynamics, disordered systems, out-of-equilibrium phenomena }
\begin{document}

\title{Comparative review of aging properties in spin glasses and \\
other disordered materials}
\author{J.Hammann, E.Vincent, V.Dupuis, M.Alba, M.Ocio and
J.-P.Bouchaud,  } 

{\scriptsize \it  To appear in the Proceedings of the workshop
``Frontiers in Magnetism'', Kyoto Oct.99}

\maketitle

\sloppy

\section{Introduction}

In materials science, aging is most commonly viewed in terms of a
steady degradation of the properties of the considered material and is
mostly a consequence of the action of some external factor (chemical
reaction, irradiation effects, mechanical fatigue ...). However many
materials and especially disordered materials experience a spontaneous
long time evolution, without any external agent, whose direct
consequences can be as important as for the usual chemical aging. This
is well known in structural glasses and amorphous polymers and is
referred to as {\it physical aging} \cite{Struik}. Much work has been
done on these systems, since it was first recognised that their
quenched state was unstable and evolved as a function of time (the
{\it aging time } \cite{Simon,Kovacs}). As quoted by L.C.E. Struik
\cite{Struik} \it{``the phenomenon of physical aging is very important
from a practical point of view. Several properties of glassy polymers,
e.g. their small strain mechanical properties, undergo marked changes
and strongly depend on the {\rm aging time}. In the testing of such
plastics, the aging time is just as important as other parameters such
as temperature, stress level, humidity, etc. Furthermore, a knowledge
of the aging behaviour of a material is indispensable to the
prediction of its long-term behaviour from short term tests''.}  \rm
This sets the real problem in the field of out-of-equilibrium
phenomena and has become a great challenge for modern statistical
physics.

In recent years, much interest has been focused on the aging properties of various disordered systems: orientational glasses, supercooled
liquids, charge density waves, and vortices in superconductors. Many
efforts have also been devoted to disordered antiferromagnets in
relation with the random field Ising model, or to
disordered ferromagnets and the problem of domain growth in presence
of randomly distributed pinning centers \cite{Young}.  Among all these examples,
spin glasses are of special interest.  They are the closest
realisation of simple theoretical models of randomly interacting
objects. There has been tremendous theoretical and experimental work
on the slow and out-of-equilibrium dynamics of spin glasses. The
question arises on how all these results compare with those of 
other types of glassy systems. Lately there has been quite an effort
to unify  experimental procedures and to generalise  theoretical
approaches. This makes a meaningful if not complete comparison
possible \cite{Young,TheoReview}. It is the aim of this paper to review some important
features of aging phenomena and discuss their relevance in
different types of systems. Two questions are on stake: Is there any
universality in the aging phenomena ? How can the comparison help
define a better understanding and a more accurate description of the
aging phenomena ?

\section{Aging time dependence and scaling properties}


\begin{figure}[htbp]
\begin{center}
\epsfysize=5.5cm\epsfbox{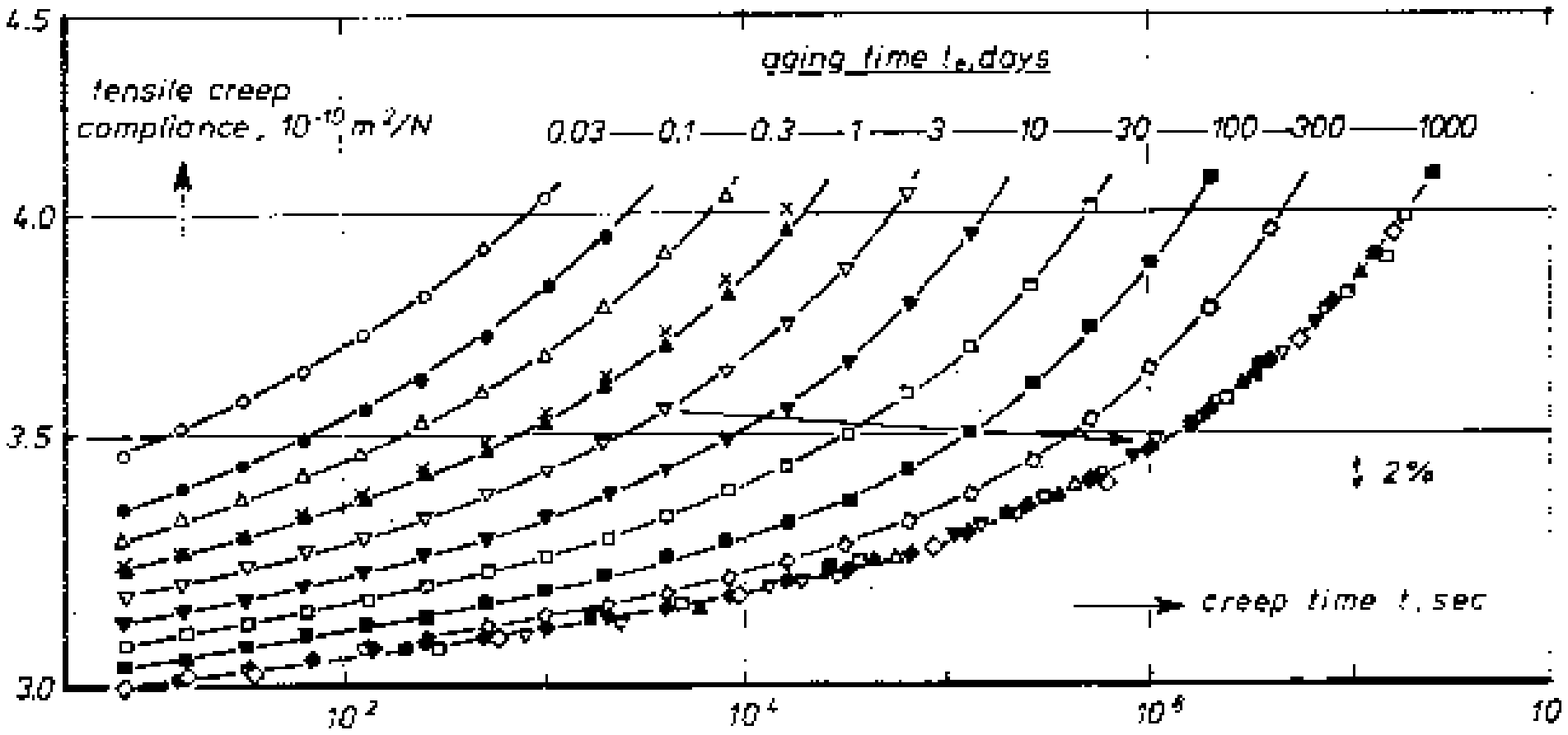}
\end{center}
\caption{\label{fig1}Response of a PVC sample submitted to a small
stress after aging during $t_e$, as a function of time (from \cite{Struik}).}

\end{figure}

Experimental evidence of aging appears, for instance, through the time
dependence of the response functions at a given temperature after an
initial quench from above the freezing temperature $T_g$. As the state of
the system evolves with the aging time $t_{a}$, which is by definition
the total time spent at the working temperature after the initial
quench, the dynamical response functions also change with
$t_{a}$. Hence, in the response to a small
excitation applied after waiting a time $t_w$ from the quench, 
two time scales are involved: $t_{a}$, the aging
time, and $t$, the observation time, i.e the time following the
application of the excitation. The aging time $t_{a}$ is
then equal to $t_{w}+t$.

Fig.1, taken from \cite{Struik}, gives a generic example of the aging
behavior in the mechanical properties of amorphous polymers. In these
experiments, a PVC sample is quenched from $90^\circ C$ (above the
freezing 
temperature $T_{g}$)  down to the working temperature $T=20^\circ C$.  A
small mechanical stress is applied after a certain waiting time $t_{w}$ (here
denoted as $t_e$) and the strain is measured as a function of time
$t$. The plot shows the very slow relaxations and their strong $t_w$
dependence up to the largest possible values of the waiting
time. Similar properties are observed for the thermoremanent
magnetisation (TRM) in spin glasses
\cite{Lundgren83,oldaging,Sitges}. Fig.2 reports results obtained on a
silver compound containing 2.6\% manganese impurities (Ag:Mn2.6\%,
$T_{g}=10.4K$). In this case, the sample was quenched down to
$T=0.87T_{g}$ in a small magnetic field which is removed after the
waiting time $t_{w}$. The figure displays the magnetisation as a
function of $t/t_w$, $t$ being the time after the removal of the field
\cite{Sitges}.

\begin{figure}[htbp]
\begin{center} 
\epsfysize=5.5cm\epsfbox{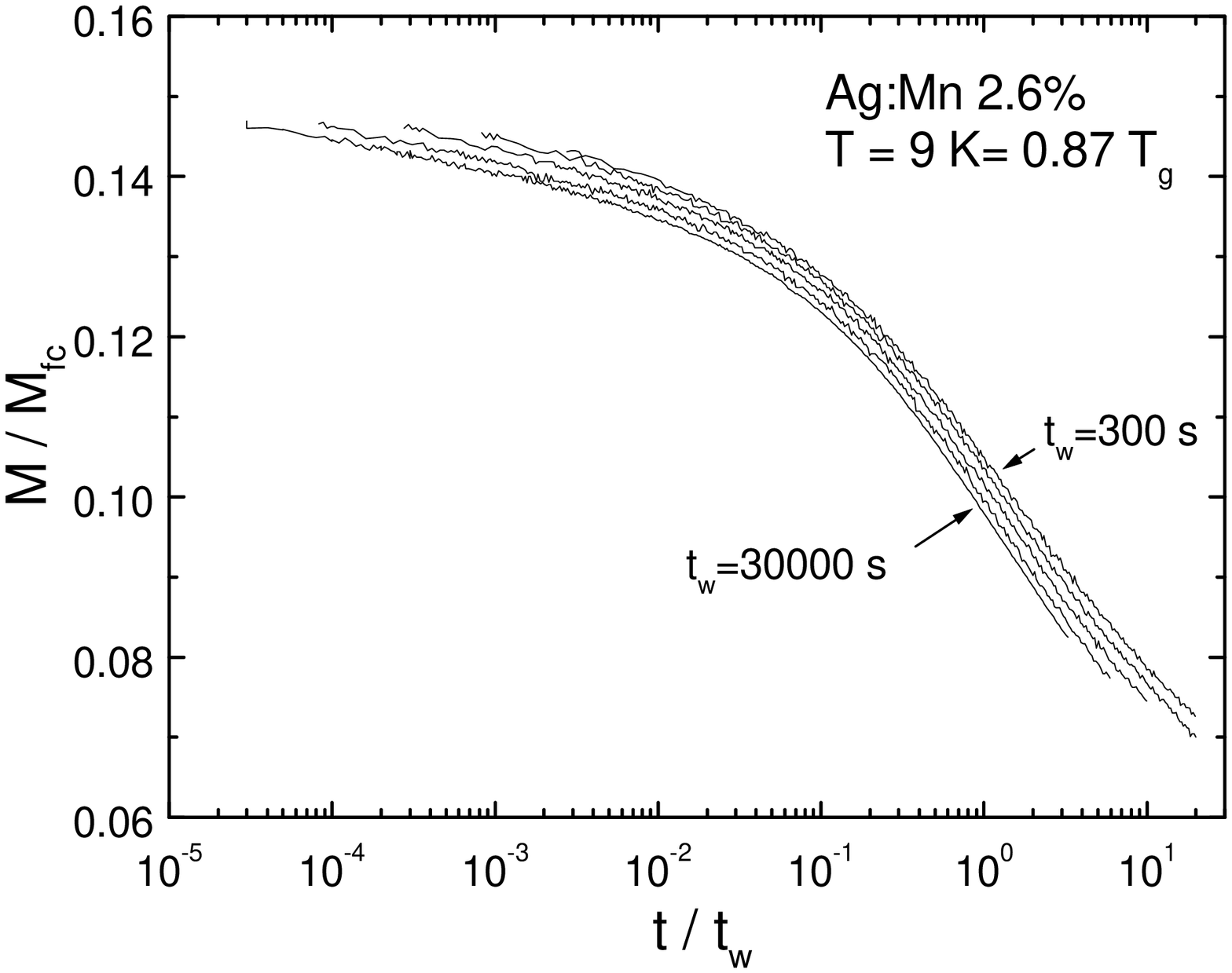}
\hspace*{0mm}
\epsfysize=5.5cm\epsfbox{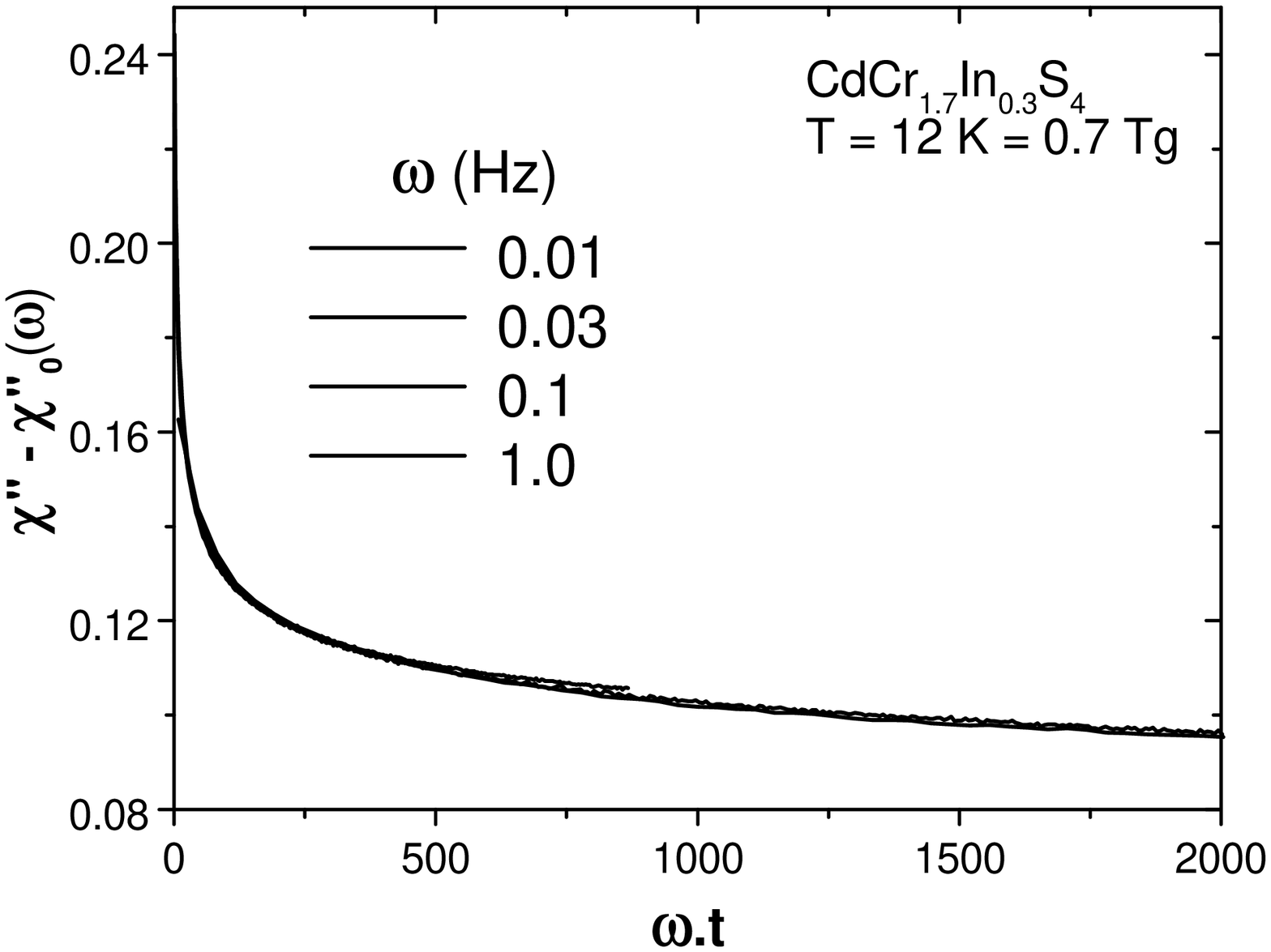}
\end{center}
\hbox to \textwidth{\vspace*{5mm}\hfil Fig.\ \ref{fig2}\hfil
\hfil Fig.\ \ref{fig3}\hfil}
\caption{\label{fig2} Ag:Mn 2.6\% spin glass, decay of the
thermo-remanent magnetisation, normalised by the field cooled value
measured just prior to removing the field (H=0.1 Oe). The $t_w$
dependence is very roughly accounted for by the $t/t_w$ scaling
variable.}
\caption{\label{fig3} 
$Cd(Cr_{0.85}In_{0.15})_{2}S_{4}$ insulating spin glass, $T=0.72T_{g}$
($T_{g}=16.7K$), relaxation of the ac out-of-phase susceptibility
$\chi''(\omega,t)$, at different frequencies $\omega$, as a function
of the time $t$ from the quench. The curves have been shifted
vertically by $\chi''_0(\omega)$, which stands for the equilibrium
(time independent) susceptibility.
}
\end{figure}

Ac measurements also give very clear evidence for the aging
phenomena. For instance, the ac shear modulus of polymers has been
investigated \cite{Cavaille}, and ac measurements of the dielectric
susceptibility are currently used in the investigation of amorphous
polymers \cite{Ciliberto}, supercooled liquids \cite{Nagel} or
orientational glasses \cite{KLT}. Fig.3 gives, in the spin glass case,
an example of the time dependence of the magnetic ac susceptibility.
The observed behavior of the out-of-phase component $\chi''$ at
frequency $\omega$ corresponds to
$\chi''(\omega,t)=\chi''_{0}(\omega)+B(\omega t)^{-\beta}$, where
$\chi ^{''}_{0}(\omega)=A\omega ^{\alpha }$ is the infinite time
``equilibrium'' value of the susceptibility, $\beta \sim 0.2-0.4$, and
$\alpha $ increases from 0.05 at low $T$ to 0.2 close to $T_{g}$.

Within the time scales used in these ac measurements, the aging part
of the magnetic susceptibility of spin glasses is seen to scale as
$\omega t$. This should imply a $t/t_{w}$ dependence for the aging
part of the thermoremanent magnetisation as long as the excitation
field is small enough to be in the linear regime for the response
function.  In the plot of fig. 2, a clear and systematic departure
from a good $t/t_w$ scaling can be seen, for two reasons: the
equilibrium relaxation (corresponding to $\chi''_{0}(\omega)$ in the
ac experiment) has not been subtracted, and also $t/t_w$ is
not the correct scaling variable (a more detailed analysis is given in
\cite{Sitges}). 
This latter observation has early been made by Struik in amorphous
polymers \cite{Struik}. It was noted that the relaxation curves of the
compliance could only be superposed  {\it in the short time
limit}, by a logarithmic shift of the time scale with a factor $\mu
=0.9$, hence a scaling of the type $t/t_{w}^{\mu}$. This suggested a
logarithmic shift of all the response times, i.e a scaling for their
distribution as $g(\tau ,t_{a})=G[\tau/(t_{a}^{\mu }\tau _{*}^{1-\mu
})]$, where $t_{a}=t_{w}+t$ is the aging time and $\tau _{*}$ some
time normalisation factor. With this assumption any infinitesimal
change of magnetisation can be written as: ${dm}/{m} = {dt}/{\tau} =
{dt}/({t_{a}^{\mu }\tau _{*}^{1-\mu })}\ \ .$ Changing the time
variable $t$ into an effective time $\lambda $ defined by $d\lambda
=dt.({t_{w}}/{t_{a}})^{\mu }$, ${dm}/{m}$ becomes: ${dm}/{m}=
{d\lambda }/ {(t_{w}^{\mu } \tau_{*}^{1-\mu } )}$.

This entails the following scaling form for the aging part of the
relaxation function: $m(t,t_{w})=M[{\lambda }/({t_{w}^{\mu }\tau
_{*}^{1-\mu }})]$. The effective time $\lambda $ is equal to $t$ for
$t\ll t_{w}$ $(t_{a}\approx t_{w})$.  For larger values of $t$, it
accounts for the change of aging time during the relaxation, which for
long $t$ is indeed not measured at a constant age of the system.  Such
a scaling for the relaxation function works remarkably well for the
amorphous polymers. It works as well for all known spin glasses. An
example is given in fig.4, where the whole set of data corresponding
to $t_{w}$'s ranging from 300s to 30000s collapse on the same master
curve \cite{oldaging,Sitges}.

\begin{figure}[htbp]
\begin{center}
\epsfysize=7cm\epsfbox{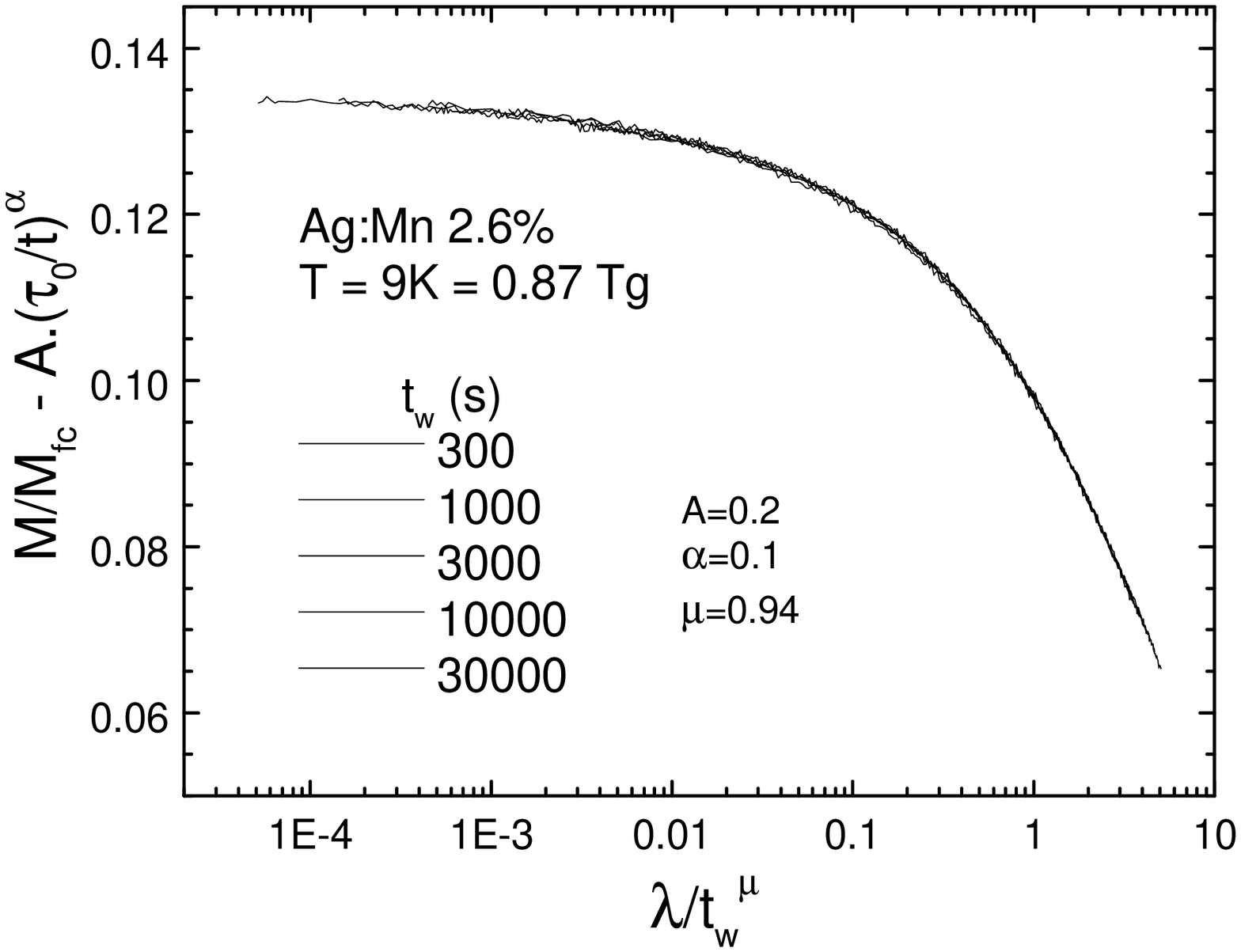}
\end{center}
\caption{\label{fig4} Aging part of the TRM (same data as in Fig.2), obtained after subtracting to the TRM a stationary contribution which is
equivalent for the magnetisation to the ``equilibrium" value $\chi''
_{0}$ for the susceptibility. 5 curves, obtained for $t_w$= 300, 1000, 3000, 10000 and 30000 s are superimposed as a function of the scaling variable defined in the text.}

\end{figure}

In other systems than spin glasses, the experimental results lead to a
whole diversity of conclusions. It was found that the ac
susceptibility of disordered ferromagnets followed an $\omega t$
scaling as in spin glasses \cite{SGferro}, but the ac dielectric
susceptibility of supercooled glycerol \cite{Nagel} and of the
orientational glass $K_{1-x}Li_xTaO_3$ with $x=0.025$ (KLT) \cite{KLT}
did not. In the case of the orientational glass, the dielectric
constant is frequency independent and only depends on the aging time
after the quench.

\section{Cooling rate effects}

In most experimental situations the system is not instantly quenched, the
cooling time is not quite negligible and may have an important
effect. This is clearly the case for the KLT orientational glass
\cite{KLT}, which indeed shows a huge dependence on the cooling rate.
The level of the in-phase dielectric susceptibility $\epsilon ^{'}$ is
quite different for the various cooling rates explored. The difference
is much larger than the amplitude of the relaxation itself, leading to
an apparent equilibrium value at ``infinite'' $t_a$ which depends on
the cooling rate. The smaller the cooling rate, the lower the value
for $\epsilon ^{'}$, and the smaller the amplitude of the
relaxation.\ {\it In 
this system, relaxations at higher temperatures strongly affect the
low temperature value of the response function.}

This is also the case in disordered
ferromagnets.  Due to  ``impurities'' (or defects) 
acting as pinning centers,  domain walls propagate  very slowly
through the network and the equilibrium distribution of domains can
only be reached in very long times. The dynamics is fastest at the
highest temperature compatible with the ferromagnetic groundstate, 
i.e. close to the transition temperature.  This is seen in Fig.5 which
shows the behavior of the magnetic susceptibility of a ferromagnetic   thiospinel 
sample around the ferromagnetic transition. This system
($CdCr_{1.9}In_{0.1}S_4$ \cite{Nogues}) is a $Cr^{3+}$ 95\%
version of the spin-glass thiospinel of Fig.3 ($Cr^{3+}$ 85\%). 
The out-of-phase
susceptibility displays a marked hysteresis between successive cooling
and heating procedures. The width of the hysteresis decreases as the
temperature sweeping rate is decreased. The downward shift of the
out-of-phase susceptibility $\chi''$ is related to the decrease of dissipative
processes as the overall surface of the walls decreases. The effect on
the in-phase susceptibility $\chi'$ is very small since this quantity is a
volume response.

\begin{figure}[htbp]
\begin{center}
\epsfysize=6cm\epsfbox{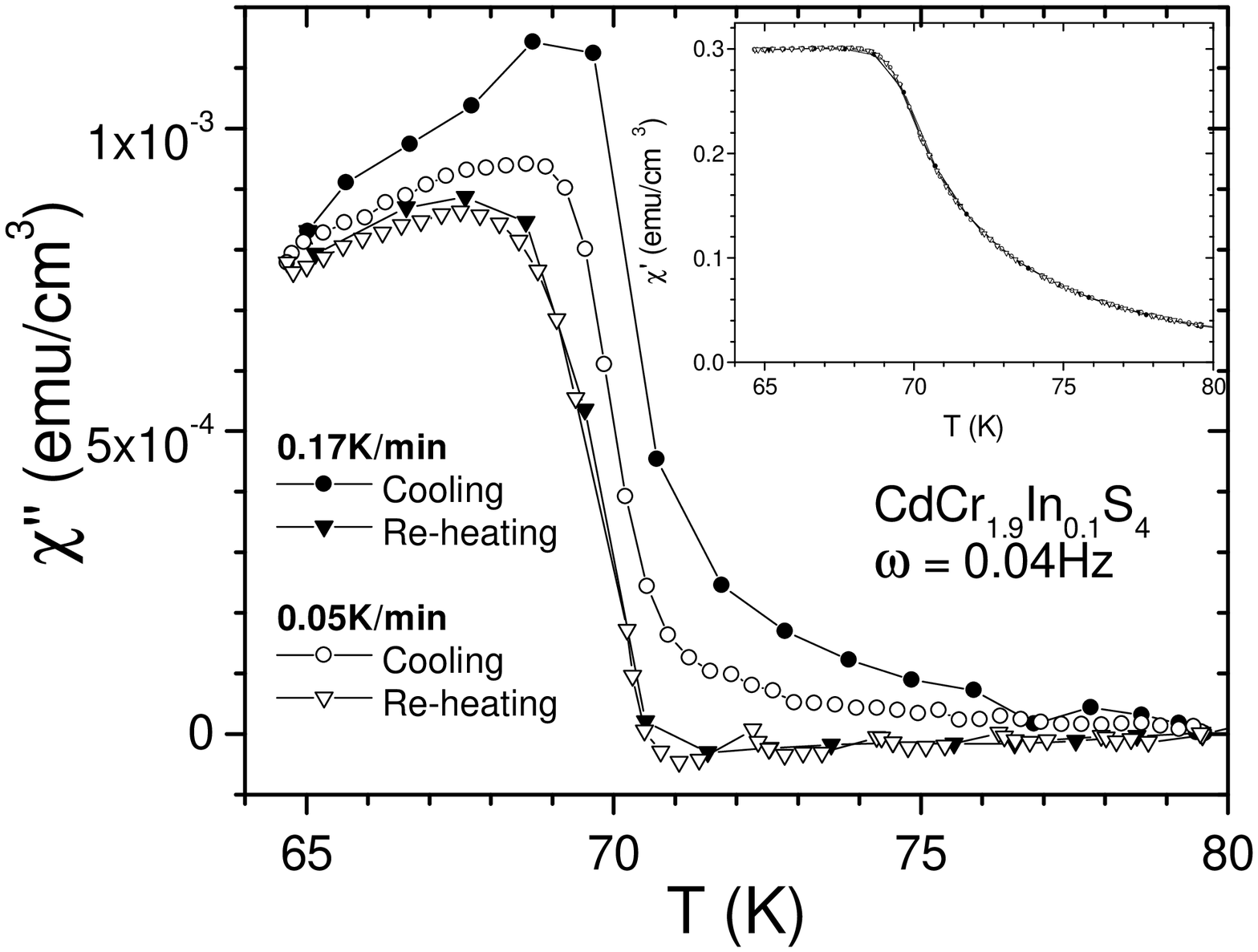}
\end{center}
\caption{\label{fig5} $CdCr_{1.9}In_{0.1}S_4$ disordered ferromagnet: hysteretic behaviour of ($\chi''$ ) in the
vicinity of the ferromagnetic transition, upon cooling and re-heating
at the different rates of $0.17K/min$ (full symbols) or $0.05K/min$
(open symbols). No hysteresis is visible on $\chi'$ (inset). }

\end{figure}

Similar cooling rate effects are present in
supercooled liquids and in amorphous polymers. For instance,
supercooled glycerol has a much lower value of the dielectric constant
if aged at an intermediate temperature, compared to its value
following a quench \cite{Nagel}. In the case of amorphous polymers, cooling rate
effects are found in temperature cycling experiments \cite{Ciliberto}. 

In this regard, spin glasses are quite singular,  cooling rate
effects are indeed very small compared to  overall aging dynamics
\cite{hystSG,memchaos}.  Fig.5 shows the influence of a very long stay
(t=900min) right below the transition ($T/T_g=0.96$) on the relaxation
at lower temperature ($T/T_g=0.72$) in the $CdCr_{1.7}In_{0.3}S_4$
thiospinel spin glass \cite{old87}.  The TRM decay,  as well as 
the $\chi''$ relaxation (inset), are completely
insensitive to the 900 min stay at 0.96 $T_g$.

It is clear that the previous thermal
history before reaching the working temperature had no relevant
effect. {\it Aging at  higher temperature did not bring the
system closer to its equilibrium at lower temperatures.}\  This
conclusion is  strongly supported by the temperature cycling
experiments.

\begin{figure}[htbp]
\begin{center}
\epsfysize=7cm\epsfbox{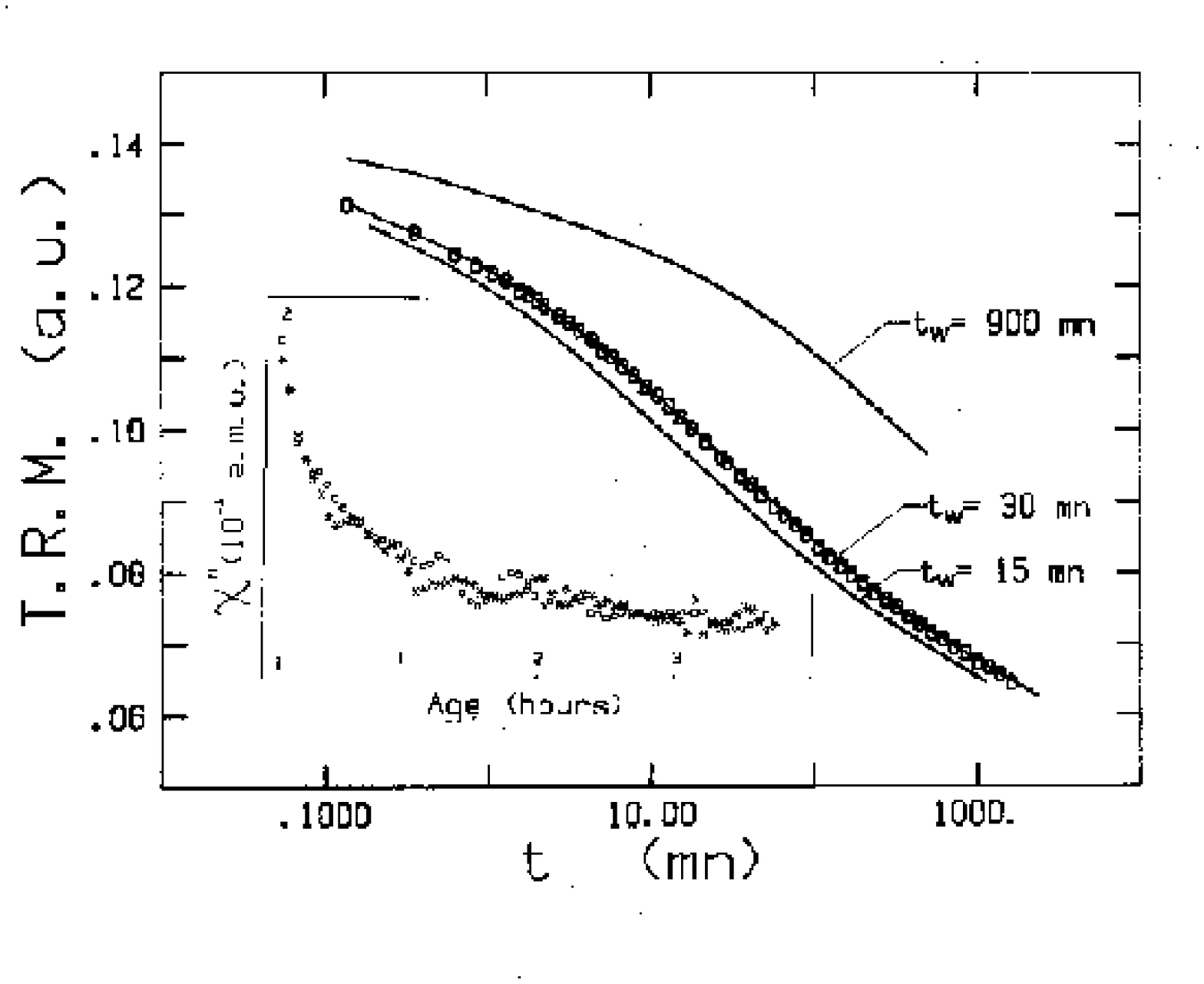}
\end{center}

\caption{\label{fig6} 
$CdCr_{1.7}In_{0.3}S_4$ thiospinel spin glass ($T_g=16.7K$): 
influence of a $900 min$ aging time at
16K ($0.96T_g$) on aging at 12K ($0.72 T_g$). 
The main part of the figure displays TRM relaxations. 
Solid lines are standard
measurements for $t_w=15,30,900$ min. Open circles: 900 min stay at
16K, and waiting time of 30 min at 12K. 
The inset shows the result of the same procedure in a 
$0.01Hz$ $\chi'' $ measurement. The stars are obtained
for a direct quench, while the circles
correspond to the intermediate waiting at 16K. 
The data are seen to be independent of the
experimental procedure (from \cite{old87}). 
}

\end{figure}

\section{Temperature cycling experiments}

Fig.7 gives an example of the spin glass behavior during a {\it negative temperature cycling procedure} \cite{Sitges,CuKuDT}.
The first remarkable point is the
upward jump of $\chi''$ and the subsequent strong relaxation after the
temperature decrease from 12K to 10K, even though the relaxation
at 12K was already rather close to its equilibrium value. The
observed relaxation at 10K is the same as the relaxation obtained
after a direct quench from above $T_g$ to 10K. {\it The aging 
processes appear to be reinitialised during a downward temperature
change ({\rm rejuvenation} effect)}. The second remarkable point is that, as T comes
back to 12K, $\chi''$ retrieves the value it had reached after the
initial aging at 12K and resumes its relaxation. The
inset of the figure shows that both relaxations at 12K are in exact
continuation of each other.
The temperature decrease and the relaxation at the lower temperature did not affect
the state of the system at 12K. {\it The system kept the whole
memory of its previous state ({\rm memory} effect)}. 
\begin{figure}[htbp]
\begin{center}
\epsfysize=6cm\epsfbox{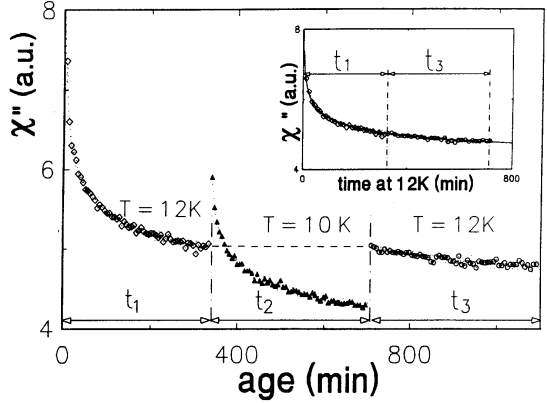}
\end{center}
\caption{\label{fig7} $CdCr_{1.7}In_{0.3}S_4$ spin glass ($T_g=16.7K$): effect of a negative temperature cycling
$12K->10K->12K$ on the time dependence of $\chi'' $ (f=0.01Hz). The inset shows the relaxation
measured during $t_3$ plotted in continuation of the initial
relaxation during $t_1$ (the solid line 
is a relaxation at $0.72T_g$ without temperature cycling).
}
\end{figure}

\begin{figure}[htbp]
\begin{center}
\epsfysize=6cm\epsfbox{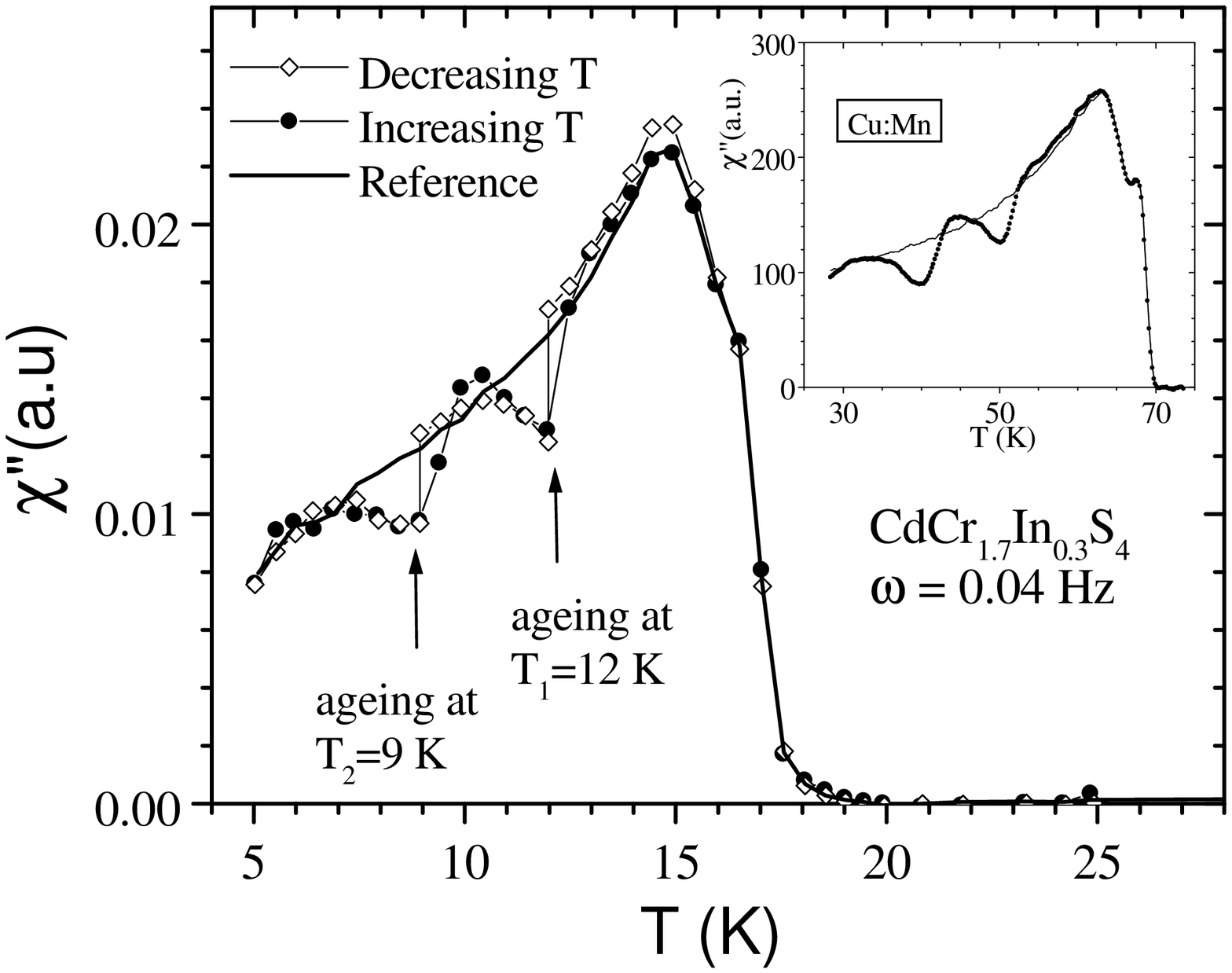}
\end{center}

\caption{\label{fig8} Rejuvenation and memory effects on the ac
out-of-phase susceptibility of the $CdCr_{1.7}In_{0.3}S_4$ spin
glass. The measuring frequency
is 0.04Hz with an ac field of 0.3 Oe and a temperature sweeping
rate of 0.1K/min. While cooling down (open diamonds), two stops are
made: one at 12K during 7h, the other one at 9K during 40h. The
reference curve (solid line) is measured during continuous re-heating
after continuous cooling. The inset shows the same result on the Cu:Mn
spin glass \cite{CuMnMC,memchaos}.}
\end{figure}

These rejuvenation and memory effects can be nicely visualized using
an alternative temperature cycling procedure \cite{memchaos}. In this
procedure (Fig.8), the susceptibility is continuously recorded as a function
of temperature.  The solid line in Fig.8 is a
reference curve, obtained in a continuous temperature sweep at
0.1K/min. In another run, the continuous cooling is stopped at 12K for
an aging stage of 7h (open diamonds),
 during which $\chi''$ relaxes downwards.
 The
rejuvenation effect appears upon resuming the cooling from 12K: the
$\chi''$ data increase back up to the reference curve within a rather
small temperature variation. Yet, this reference curve is not an
equilibrium curve; when, at 9K, the cooling is stopped once more (during
40h), $\chi''$ relaxes strongly downwards, 
and again merges with the reference curve when cooling continues down to
5K.   
When the sample is now continuously re-heated,
$\chi''$ displays two sharp dips centered at 9 and 12K, matching the
values reached after the aging period. Aging at both temperatures of
9 and 12K has remained imprinted in the system during the whole run
at lower temperatures.  As soon as the temperature becomes higher than
that at which aging occurred, the memory effect disappears, and any
further cooling procedure simply follows the normal reference
curve. The inset of Fig.8 shows that the behavior observed for the
$Cr^{3+}$ 85\% thiospinel insulating compound is found with exactly the same
characteristic features in the metallic spin glass Cu:Mn
\cite{CuMnMC}.

Similar experiments have been performed \cite{SGferro} on the
$Cr^{3+}$ 95\% thiospinel sample of Fig.5 ($CdCr_{1.9}In_{0.1}S_4$
\cite{Nogues}), which shows a ferromagnetic transition at $T_c=70K$,
followed at $T_g=10K$ by the reentrance of a spin-glass phase. At all
temperatures below $T_c$, ferromagnetism has been characterized by
neutron diffraction measurements \cite{neutrons}. Important
irreversibilities are evidenced below $T_c$ by a large split between
the zero-field cooled and field-cooled curves, indicative of slow
dynamics. Fig.9 presents the out-of-phase susceptibility $\chi''$ of this
sample close to $T_c$ as a function of temperature, using a procedure equivalent to
that of Fig.8. The curve with open diamonds corresponds to a run in
which the temperature was decreased down to $T_1=67K$ where the system
was aged for a time of 3.5h. This aging leads to a noticeable decrease
of $\chi''$. The temperature was then further decreased, down to
temperatures $T_d$ ranging between 64 and 30K.  The curve never comes
back exactly to the reference curve, but there is clearly some
rejuvenation, since $\chi''$ first increases before decreasing again
as the temperature is lowered from $T_1$. In the subsequent reheating
procedure, no dip can be observed in the case of the lowest $T_d=30K$
(open circles). Thus no memory effect exists in this case. However, if
the lowest temperature $T_d$ reached in the run comes closer to $T_1$,
a partial memory can be detected. The amplitude of the memory effect
increases as $T_d$ becomes closer and closer to $T_1$. In the
reentrant spin-glass phase below 10K, the same rejuvenation and (full)
memory effects as in usual spin glasses have been found\cite{SGferro}.

Very recently, the non-diluted thiospinel $CdCr_2S_4$ has also been
studied. Below the ferromagnetic transition at $T_c=85K$, important
irreversibilities are again found. At low temperature (5K),
surprisingly, $\chi''$ shows a peak which is similar to that observed
at the onset of the reentrant phase in the diluted
thiospinels. However, the $\chi''$ relaxations in this region are very
weak compared to the spin-glass case. In these thiospinel compounds,
the first-neighbour interactions are ferromagnetic, while the next
nearest-neighbour interactions are antiferromagnetic. It is therefore
not excluded \cite{Miyashita} that the $\chi''$ peak signs up the
reentrance of a strongly disordered (but almost frozen)
low-temperature state, even in this non-diluted compound. Slightly
below $T_c=85K$, the same experimental procedure as in Fig.9 has been
applied. In this region close to the ferromagnetic transition, aging
relaxations are important, and the same rejuvenation and (partial)
memory effects are found as in the diluted compound of Fig.9.

Similar experimental procedures have been applied to other systems. In
most cases \cite{Ciliberto,KLT,Nagel}, no
strong rejuvenation effects are found when going to lower
temperatures. Aging at different temperatures is essentially
cumulative, and is mainly important in the vicinity of the glass
transition \cite{Ciliberto}. During a negative temperature cycling, all dynamics is
considerably slowed down, and a ``memory'' of the previous stage is
then naturally retrieved when heating back. However, this memory can
be affected by the low-temperature evolution \cite{KLT,Cavaille,Nagel}, yielding a non-monotonic
behavior of the susceptibility after the
negative cycling (this is denoted as a ``memory effect'' in the case
of polymers \cite{Cavaille}). In the dielectric
orientational glass $KTa_{0.97}Nb_{0.03}O_3$ (KTN \cite{KTN}), however,
rejuvenation and partial memory effects are found, in a very similar
way as observed in the disordered ferromagnet Fig.9.

\begin{figure}[htbp]
\begin{center}
\epsfysize=6cm\epsfbox{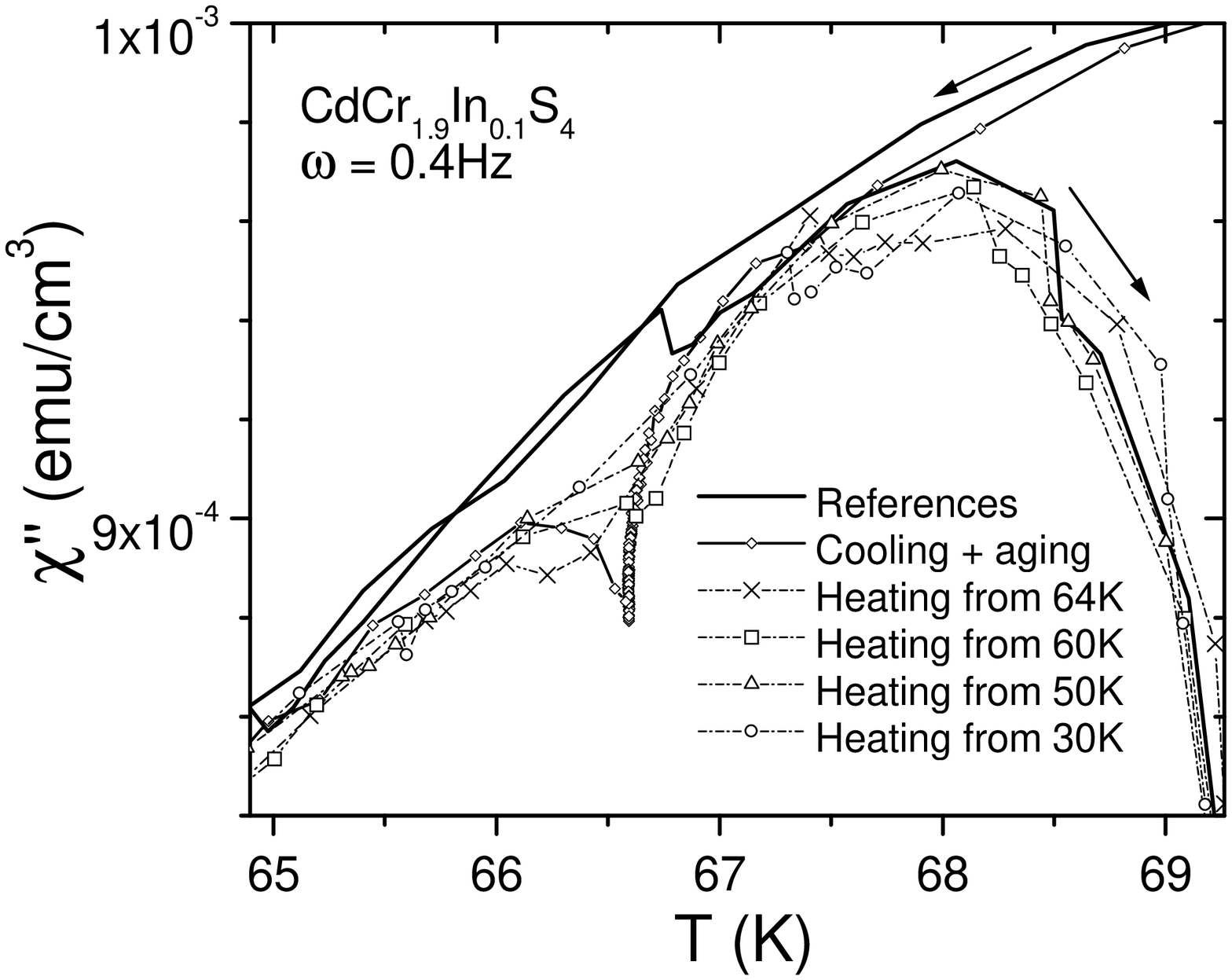}
\end{center}

\caption{\label{fig9} Rejuvenation and partial memory effect in the
disordered ferromagnet $CdCr_{1.9}In_{0.1}S_4$.
The procedure is similar to
that of Fig.8, except that the low-temperature excursion was limited
to either $T_d=64K$ (crosses), $60K$ (squares), $50K$ (triangles) or
$30K$ (circles). For a short enough low-temperature excursion (e.g.
$T_d=64K$, crosses), the re-heating curves show a partial memory
effect, which progressively faints out for lower and lower $T_d$
values. }

\end{figure}

\section{Summary}

{\it A universal feature of the aging properties of disordered materials
seems to be the existence of two intricated types of dynamic
processes, whose relative importance changes from one system to the
other.} On the one hand, {\it cumulative} processes bring the system towards
the same ground state whatever the temperature. The time evolutions
at different temperatures are adding up, yielding important cooling
rate effects. On the other hand, {\it temperature specific} processes
(rejuvenation and memory effects) correspond to a restart of aging at
each different temperature, but the previous relaxations at higher
temperatures remain imprinted, and can be retrieved as the system is
heated back to these temperatures. In polymers \cite{Ciliberto}, in
supercooled glycerol \cite{Nagel} and in the orientational glass KLT
\cite{KLT}, the cumulative part was the most important. In disordered
ferromagnets \cite{SGferro} and in KTN \cite{KTN}, both contributions
appear almost equally.  In spin glasses, this cumulative part is
almost negligible: the spin glass dynamics is mostly characterized by
the rejuvenation and memory effects.

\section{Analysis and conclusion}

With reference to the ferromagnetic behavior, the cumulative part of
the dynamics may be ascribed to slow domain coarsening in a medium
with randomly distributed pinning centers. The second (non-cumulative)
part could then be due to domain wall reconformations, the domain
walls trying to find the most favourable path through the nearby
pinning centers with the smallest loss of elastic energy. Such an
interplay between  elastic forces acting in walls and random local
pinning forces has been shown to lead to a hierarchical free energy
structure \cite{Balents}. The reconformations indeed involve
rearrangements over a large spectrum of characteristic length scales,
the small length scales being affected by the large ones. Such
reconformations can occur even at almost constant domain sizes.

It was already clearly demonstrated that a hierarchical free energy
structure well accounts for the rejuvenation and memory
effects\cite{hierarchie}. These can also be directly
understood within the scheme of wall reconformations in the following
way. Each rearrangement over a characteristic length $l$ is associated
to an energy barrier $E(l)$ increasing with $l$. This energy is itself
associated to a characteristic time through thermal activation. A
given temperature $T$ selects out a length $l^*$ such that $E(l^*)\sim
k_BT$. Then all the modes corresponding to shorter length scales,
hence smaller barriers, will be equilibrated in short times, whereas
modes with larger length scales will correspond to much longer times
and thus experience slow aging or almost complete freezing for the
largest lengths \cite{JPBrevue}. This accounts for the coexistence of
equilibrium (stationary) and aging (non-stationary) dynamics. As $T$
is decreased, $l^*$ decreases. The slow modes freeze out, while the
fast modes slow down and fall out of equilibrium due to the change of
Boltzman weights. Some of these now enter the experimental time window
and are seen as restarting aging processes (rejuvenation effect). As
$T$ is brought back to the previous higher value, the frozen modes
become active again and continue their former aging, while the other
modes again become fast and equilibrate in short times (memory
effect). If a significant domain coarsening has occurred during the
temperature cycling, the wall reconformations take place in a
different pinning environment after the end of the cycle, and the
previous rearrangements are not relevant any more. The memory of aging
at the initial temperature is lost.

The implication of this picture for spin glasses would be that domain
coarsening is almost irrelevant, whereas the predominant part of
dynamics is related to wall reconformations. The spin glass phase
would thus appear as a dense network of domain walls
\cite{SGferro,Martin,JPBrevue,fractdiv}. In the absence of
ferromagnetism, the exact nature of such walls and of the domains
themselves (are they made of only two or many phases?) remains an open
question.  The growth of a spin-spin correlation length during aging
at constant temperature has been characterized in simulations
\cite{numerics} and in experiments \cite{OrbachKsi}. If spin-glass
dynamics is to be seen as {\it wall dynamics}, the observed growth of
a correlation length should be related to a progressive increase of
the largest length scales involved in the domain wall
reconformations. Rejuvenation effects probably imply that other
(smaller) length scales are simultaneously involved.


\begin{thebibliography}{}
\bibitem{hystSG} In the thiospinel sample of Fig.8, the
$\chi''$-hysteresis between continuous cooling and heating curves is of $\sim
6\%$ at $0.7T_g$, whereas aging  during 7h at the same temperature
amounts to $\sim 38\%$ of the equilibrium value.
\end{thebibliography}
\end{document}